\documentclass[fleqn,usenatbib]{mnras}

\usepackage{newtxtext,newtxmath}

\usepackage[T1]{fontenc}
\usepackage{ae,aecompl}

\usepackage{graphicx}	
\usepackage{amsmath}	
\usepackage{amssymb}	

\usepackage{xspace}     
\usepackage{xcolor}     
\usepackage{xstring}
\usepackage{ifthen}

\usepackage{comment}
\usepackage{blindtext}

\newcommand{\xGASS}{xGASS\xspace}	
\newcommand{\ProFit}{\textsc{ProFit}\xspace}	
\newcommand{\GALEX}{\textit{GALEX}\xspace}	
\newcommand{\SDSS}{SDSS\xspace}	

\newcommand{\Mstar}{$M_{\star}$\xspace}    
\newcommand{\MstarT}{$M_{\star,Total}$\xspace}    
\newcommand{\MstarD}{$M_{\star,Disc}$\xspace}    

\newcommand{\SFR}{$SFR$\xspace}    
\newcommand{\sSFR}{$sSFR$\xspace}    
\newcommand{\sSFRT}{$sSFR_\mathrm{Total}$\xspace}    
\newcommand{\sSFRD}{$sSFR_\mathrm{Disc}$\xspace}    

\newcommand{\HI}{H$\,\textsc{i}$\xspace}    

\newcommand{\rband}{$r$-band\xspace}            
\newcommand{\iband}{$i$-band\xspace}            

\newcommand{\gi}{$g-i$\xspace}          
\newcommand{\NUVr}{NUV$\,-\,r$\xspace}              

\newcommand{\BtoTM}{$\mathrm{B/T}_\mathrm{M_{\star}}$\xspace}          

\newcommand{\Msun}{$M_{\odot}$\xspace}    

\newcommand{\SFRvM}{$SFR$--$M_\mathrm{\star}$\xspace} 
\newcommand{\sSFRvM}{$sSFR$--$M_\mathrm{\star}$\xspace} 
\newcommand{\SFMS}{SFMS\xspace} 

\newcommand{\nSigma}{2}

\newcommand{\Sersic}{S\'{e}rsic\xspace}    

\newcommand{\perc}[1] {$#1$\,\%}
\newcommand{\simperc}[1] {$\sim#1$\,\%}
\newcommand{\supth}[1] {$#1^{th}$\xspace}

\title[Morphology in the Star-Forming Main Sequence]{xGASS: The Role of Bulges Along and Across the Local Star-Forming Main Sequence}
\author[R. H. W. Cook et al.]{
Robin H. W. Cook$^{1,2}$\thanks{E-mail: robin.cook@icrar.org},
Luca Cortese$^{1,2}$,
Barbara Catinella$^{1,2}$,
Aaron Robotham$^{1,2}$
\\
$^{1}$International Centre for Radio Astronomy Research (ICRAR), University of Western Australia, Crawley, WA 6009, Australia\\
$^{2}$Australian Research Council, Centre of Excellence for All Sky Astrophysics in 3 Dimensions (ASTRO 3D), Australia
}

\date{Accepted XXX. Received YYY; in original form ZZZ}

\pubyear{2020}

\begin{document}

\maketitle


\begin{abstract}
We use our catalogue of structural decomposition measurements for the extended GALEX Arecibo SDSS Survey (xGASS) to study the role of bulges both along and across the galaxy star-forming main sequence (\SFMS). We show that the slope in the \sSFRvM relation flattens by $\sim$0.1 dex per decade in \Mstar when re-normalising \sSFR by disc stellar mass  instead of total stellar mass. However, recasting the \sSFRvM relation into the framework of only disc-specific quantities shows that a residual trend remains against disc stellar mass with equivalent slope and comparable scatter to that of the total galaxy relation. This suggests that the residual declining slope of the \SFMS is intrinsic to the disc components of galaxies. We further investigate the distribution of bulge-to-total ratios ($B/T$) as a function of distance from the \SFMS ($\Delta SFR_{MS}$). At all stellar masses, the average $B/T$ of local galaxies decreases monotonically with increasing $\Delta SFR_{MS}$. Contrary to previous works, we find that the upper-envelope of the \SFMS is not dominated by objects with a significant bulge component. This rules out a scenario in which, in the local Universe, objects with increased star formation activity are simultaneously experiencing a significant bulge growth. We suggest that much of the discrepancies between different works studying the role of bulges originates from differences in the methodology of structurally decomposing galaxies.
\end{abstract}


\section{Introduction}
The observed correlation between a galaxy's star formation rate (SFR) and stellar mass likely contains fundamental information from which we can begin to understand the evolution of galaxies over cosmic time \citep{Brinchmann2004, Daddi2007, Rodighiero2010, Wuyts2011, Noeske2007}. From this, the star-forming main sequence (\SFMS) has become a powerful tool for understanding the origins of the distribution and evolution of galaxy properties throughout the Universe. The \SFMS is commonly parameterised via a linear relation between $\log\,SFR$ and $\log\,M_{\star}$ (i.e. a power law), with an observed scatter of $\sim 0.3$\,dex \citep{Speagle2014,Whitaker2015} and a slope ranging from 0.5 to 1.0 dex. Uncertainties in the slope arise predominantly from inconsistencies in individual \SFR calibrations \citep{Pannella2009, Davies2016} and how ``star-forming'' galaxies are defined \citep{Salim2007}. The \SFMS seems to hold over at least the last 10\,Gyr \citep{Elbaz2007, Peng2010b, Whitaker2012, Whitaker2014,Popesso2019b} with a normalisation that is observed to increase at earlier epochs \citep{Schreiber2015}. This likely reflects the sharp decline in the cosmic star formation history by a factor of 10 since $z \sim 1$ \citep{Lilly1996,Madau1998,Hopkins2006b}.

Many studies have shown that the \SFRvM relation has a slope that is less than unity (e.g. \citealt{Whitaker2014, Lee2015, Schreiber2015, Tomczak2016}). This departure from unity is most notable when the \SFMS is recast in terms of the ratio of current \SFR to current stellar mass, or specific \SFR ($sSFR \equiv SFR/M_{\star}$). This can be considered to be a galaxy's fractional mass-growth rate or, its inverse, the galaxy build-up time. Given that the \sSFRvM relation is observed to have a range of negative slopes, this implies that not all galaxies form stars at a constant efficiency throughout their evolution and that a residual mass trend may hint at the possible physical process(es) responsible for the suppression of \SFR towards higher masses. This coincides with the term ``downsizing'', which has often been used to describe the observation that more massive galaxies have formed earlier and at a faster rate \citep{Neistein2006}. Historically, the concept of \sSFR is synonymous to the birthrate parameter \citep{Kennicutt1994,Boselli2001}, expressed as the ratio of the current \SFR to the average \SFR integrated over its lifetime. Early-type galaxies generally have a small birthrate parameter indicating that most of their stars have formed at an earlier epoch.

Furthermore, there is evidence to suggest that the \SFRvM relation is not strictly a power law, but instead shows curvature in the high stellar mass ($M_{\star} \gtrsim 10.5\;M_{\odot}$) regime \citep{Elbaz2011,Whitaker2014,Gavazzi2015,Lee2015,Schreiber2015,Popesso2019a}. Such studies find a low mass power law of slope $\alpha \sim 1$, which becomes shallower above a turnover mass that ranges from $10^{9.5}$ to $10^{10.8}$\;\Msun, with evidence suggesting that this turnover mass may increase with redshift \citep{Tomczak2016}. However, other studies do not observe a mass-dependent slope (e.g. \citealt{Rodighiero2014,Speagle2014}). In addition to fitting a power law, the \SFMS can also be defined by tracing the ridge along the locus of the star-forming galaxy distribution \citep{Renzini+Peng2015}.

A suggested explanation for the existence of a flattening in the \SFMS has been attributed to the decreasing contribution of star-forming discs towards higher stellar masses. If it is assumed that the bulk of star formation occurs in the disc, then a flat, linear relation in the star-forming main sequence could remain if one considers the disc mass alone \citep{Guo2015}. \citet{Abramson2014} show that by re-normalising the \sSFR by the disc (instead of total) stellar mass (\MstarD), one can account for $\sim0.25$\,dex of declining \sSFR per decade of \Mstar.
They suggest that the discs maintain a constant \sSFR if one correctly accounts for the mass present in passive bulges. However, this result has not been confirmed by other authors who show that a constant \sSFR does not necessarily exist amongst discs considered independently \citep{Guo2015,Whitaker2015,Schreiber2015,Morselli2017}. \citet{Popesso2019a} suggest that, because the discs of high-mass galaxies are redder than their lower mass counterparts, the bending of the SFMS is instead due to the starvation of cold gas in a hot halo environment. As well as looking at the role of bulges in regulating the shape of the \SFMS, many studies have also studied how the position of a galaxy across the \SFMS relates to the growth of their central component. \citet{Morselli2017,Popesso2019a} find that the average $B/T$ of galaxies increases both above and below the \SFMS and is suggested to correspond to a central enhancement of star formation activity observed in starburst galaxies \citep{Morselli2019,Belfiore2018,Ellison2018a}. These observations have pointed towards a possible scenario in which star-forming galaxies may oscillate about the \SFMS due to successive compaction events followed by depletion of their cold gas reservoirs \citep{Zolotov2015,Tacchella2016}.

We investigate the nature of these findings in this paper using the structural decomposition of the \xGASS sample \citep{Catinella2010,Catinella2018}. Although this sample only contains $\sim$1,200 galaxies, in the context of measuring structural parameters through modelling of galaxy light profiles, it is important to note that to achieve reliable measurements of galaxy structure, large number statistics alone is not sufficient. In \citet{Cook2019}, we showed that informed model validation (beyond goodness-of-fit metrics) is necessary to consistently derive physically meaningful solutions for galaxy models. As we will show, this has important implications when understanding how morphology is linked to the evolution of galaxies.

This paper is organised as follows. In Section \ref{sec:sample}, we describe the sample used and the structural decomposition catalogue. Section \ref{sec:results} presents our results of analysing the role of structure along and across the \SFMS, followed by an analysis of the implications of poorly fit models in Section \ref{sec:bd_decomp}. In Section \ref{sec:discussion}, we discuss our results in regards to previous works and conclude in Section \ref{sec:conclusion}. All distance-dependent quantities are computed assuming $\Omega_{M} = 0.3$, $\Omega_{\Lambda} = 0.7$ and $H_{0} = 70$\,km\,s$^{-1}$\,Mpc$^{-1}$.

\section{Sample and Summary of Data}
\label{sec:sample}
The sample used throughout this paper is the extended \textit{GALEX} Arecibo \SDSS Survey (\xGASS; \citealt{Catinella2010,Catinella2018}). This survey contains 1179 galaxies selected only by redshift ($0.01 < z < 0.05$) and stellar mass ($10^{9}\,< M_{\star} <\,10^{11.5}$\,\Msun), and currently represents the deepest sample of cold gas observations for galaxies in the local Universe to date. In addition to these cold gas observations is a substantial amount of auxiliary data yielding optical and star formation properties across the full sample. The parent sample comes from the Sloan Digital Sky Survey (\SDSS) DR6 \citep{Adelman-McCarthy2008} for which \SDSS spectroscopy and \textit{Galaxy Evolution Explorer} (\GALEX) Medium Imaging Survey \citep{Martin2005} observations were available. The final sample was selected randomly such that a near flat distribution in stellar mass was achieved (see \citealt{Catinella2018} for more details). Optical parameters (excluding those derived from model fitting) were taken from the \SDSS DR7 \citep{Abazajian2009} database whilst UV properties and star formation rates (SFR) are calculated by combining near-UV (NUV) photometry (from various GALEX catalogues) and mid-infrared (MIR) photometry from the Wide-field Infrared Survey Explorer (WISE, \citealt{Wright2010}) as detailed in \citet{Janowiecki2017}. In the cases where unflagged measurements were not available in both NUV and MIR, SFRs were instead determined from the spectroscopic energy distribution fits in \citet{Wang2011}.

In \citet{Cook2019}, we present the accompanying catalogue of bulge-disc decompositions for the \xGASS representative sample using \SDSS $g$, $r$ and \iband imaging data. This catalogue aimed to robustly measure the morphological parameters of the bulges and discs of galaxies with a low level of spurious fits. We improved on previous bulge-disc decomposition catalogues with the combined usage of \ProFit \citep{Robotham2017} --- a Bayesian two-dimensional galaxy photometric profile fitting code --- with additional model filtering and constrained remodelling. Importantly, each galaxy was individually verified and wherever a poor fit was attained (e.g. due to galaxy interactions, prominent secondary features such as bars, etc.), the galaxy was remodelled with informed constraints to further discourage these issues from recurring. This was possible for a total of 1073 ($\sim$91\%) galaxies, a considerably larger fraction than would be the case without model validation. The resulting catalogue is far less contaminated by spurious model fits, which are commonplace amongst much larger samples. Whilst the additional need for constrained remodelling does not scale up to larger galaxy surveys, the added gain of minimising incorrectly modelled and/or misclassified galaxies exceeds the increased statistics one might depend upon to wash out any issues faced during the model fitting stages.

A key quantity derived from the model fitting is the bulge-to-total ratio ($B/T$), expressed as the fraction of light (or otherwise, stellar mass) contained within the bulges of galaxies in a given band. A significant fraction of \xGASS galaxies are best modelled as single-component systems, corresponding to $B/T = 0$ for the 292 pure-disc galaxies and, conversely, $B/T = 1$ for the 55 pure-bulge galaxies classified in the sample. Finally, stellar masses were determined using empirical recipes following \citet{Zibetti2009}, where we have used \rband flux measurements and (\gi) colours from the individual component profiles.

\begin{figure*}
    \centering
    \includegraphics[width=0.95\textwidth]{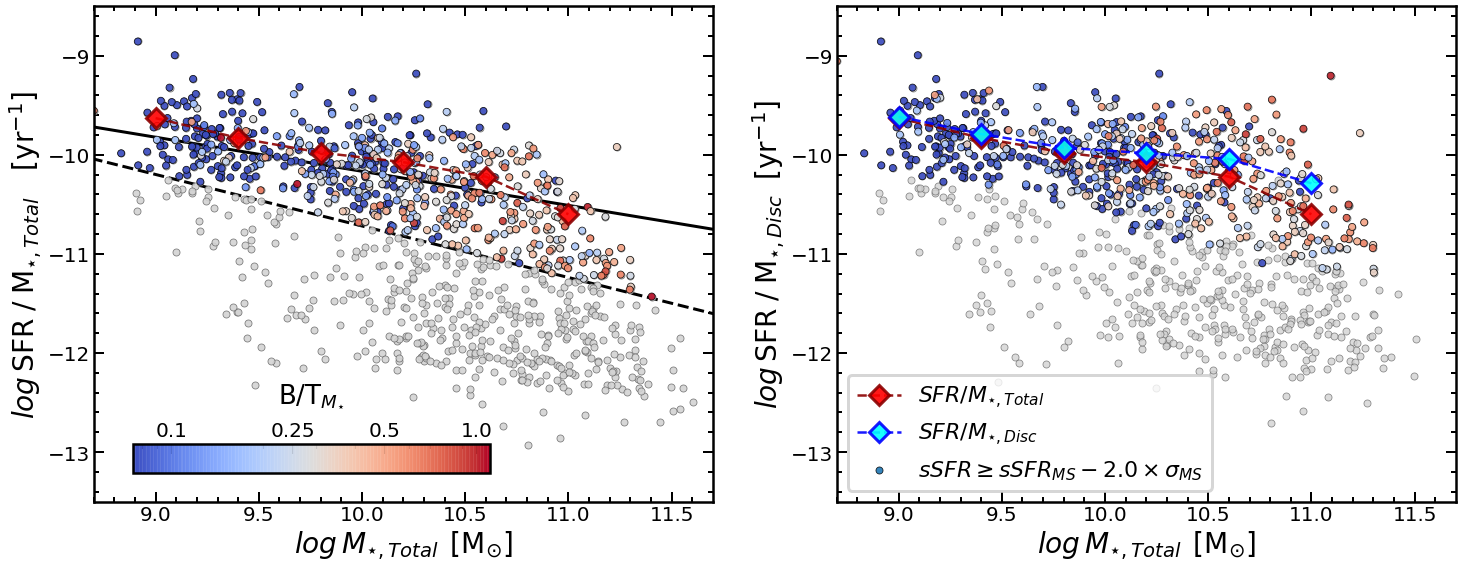}
    \caption{The main-sequence of star-forming galaxies using the specific star formation rate normalised by the total stellar mass (\sSFRT, left) or the disc stellar mass (\sSFRD, right). Smaller points show all galaxies in the \xGASS sample, with coloured points showing the subset of ``star-forming'' galaxies defined as being $\geqslant$ \nSigma\,$\sigma_{MS}$ away from the \SFMS. The solid black line represents the \SFMS (as defined in equation \ref{eqn:SFMS}) with the dashed line showing the cut $\nSigma\,\sigma_{MS}$ below. The colour of the points indicate each galaxy's stellar mass bulge-to-total ratio (\BtoTM). The diamond points show the median values of star-forming galaxies in bins of stellar mass. Grey points are not considered star-forming here and hence are not included in calculating the median. The right panel shows the medians for both the \sSFRD (blue) as well as the \sSFRT (red) for reference.}
    \label{fig:sSFR_vs_MstarT}
\end{figure*}

\begin{figure}
    \centering
    \includegraphics[width=\columnwidth]{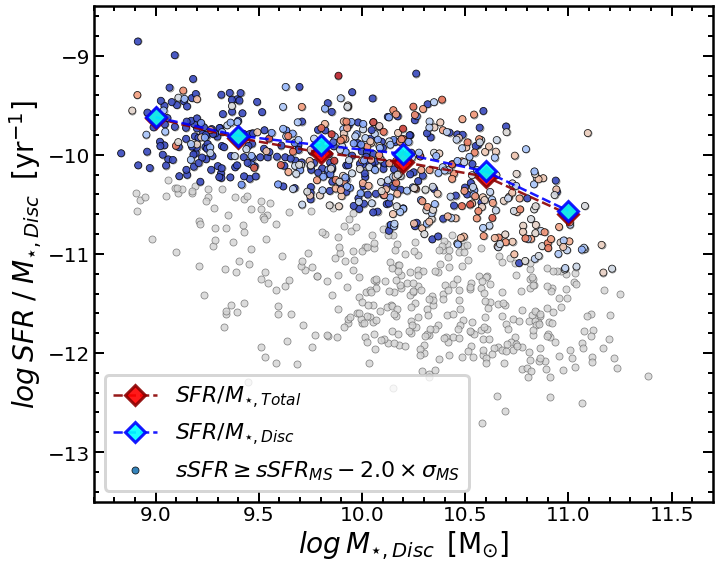}
    \caption{Same as Figure \ref{fig:sSFR_vs_MstarT} (right panel) but plotted as a function of disc stellar mass.}
    \label{fig:sSFR_vs_MstarD}
\end{figure}

\section{Results}
\label{sec:results}

\begin{figure*}
    \centering
    \includegraphics[width=\textwidth]{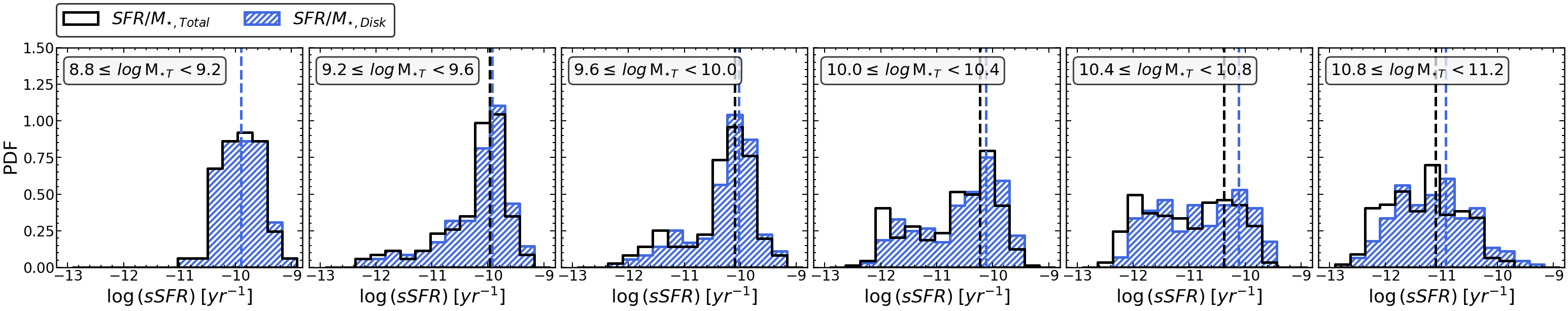}
    \includegraphics[width=\textwidth]{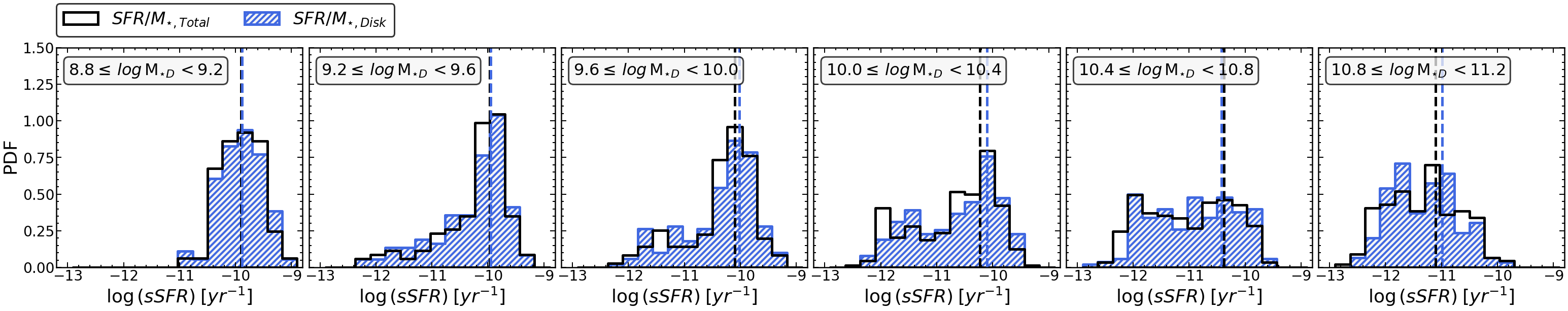}
    \caption{Top: Probability density functions of specific star formation rates in $0.4\;dex$ bins of \emph{total} stellar mass (\MstarT). The distributions of \sSFRT and \sSFRD are represented by the black and hatched blue histograms, respectively. For comparisons between \sSFRT and \sSFRD, the vertical lines show the means obtained for each Gaussian model. Bottom: repeating the analysis instead binning galaxies according to their \emph{disc} stellar mass (\MstarD). The black histogram is the same in the top and bottom panels.}
    \label{fig:sSFR_hists_by_Mstar}
\end{figure*}

\subsection{Role of Bulges Along the Star-Forming Main Sequence}
\label{sec:SFR_fixed_disk}
We begin by investigating the positions of galaxies along the $\log\,sSFR - \log\,M_{\star}$ plane to understand the origin of the residual dependence on mass in the slope of the \SFMS. In particular, whether this can be explained by the morphological transition that occurs towards higher stellar masses, where bulges (themselves typically not star-forming) become increasingly predominant. The left panel of Figure \ref{fig:sSFR_vs_MstarT} shows the \sSFRvM plane for all \xGASS galaxies with the subset of \emph{star-forming} galaxies shown as coloured points. Here, star-forming galaxies have been defined as those with a \SFR greater than \nSigma\,$\sigma$ below the \SFMS as defined in \citet{Catinella2018}; see also \citet{Janowiecki2019}. The \xGASS \SFMS is given by the following expression:

\begin{equation}
    {\rm log}\,(sSFR_\mathrm{MS}) = -0.344\left({\rm log}\,(M_{\star}) - 9\right) - 9.822
    \label{eqn:SFMS}
\end{equation}

\noindent
with a corresponding scatter ($\sigma_{MS}$) as a function of stellar mass given by:
\begin{equation}
    \sigma_\mathrm{MS} = 0.088\left({\rm log}\,(M_{\star}) - 9\right) + 0.188,
    \label{eqn:sigmaSFMS}
\end{equation}

\noindent
which, for the range of stellar masses in \xGASS, equates to a typical scatter in the \SFMS of $\sigma_{MS} = 0.2$\,--\,$0.35$\;dex; in agreement with many previous studies \citep{Daddi2007,Speagle2014,Popesso2019a} at this redshift.

The right panel of Figure \ref{fig:sSFR_vs_MstarT} shows instead the re-normalisation of the specific \SFR by the disc stellar mass, defined here as:

\begin{equation}
    sSFR_{Disc} \equiv SFR/M_{\star,Disc}.
    \label{eqn:sSFRD}
\end{equation}

\noindent
This quantity reflects the fractional mass-growth of the disc alone. Whilst there may be some contribution to the \SFR from a bulge or nuclear region, this is likely quite small. In both panels, star-forming galaxies are coloured according to their stellar mass bulge-to-total ratio, hence indicating by how much each galaxy will shift when plotted by its disc mass. Blue points --- being pure-disc systems --- do not move as, by definition, all of their mass is contained within their disc. The larger diamond points show the median $\rm{log}$\,\sSFR for star-forming galaxies within each \Mstar bin. There is indeed a slight flattening ($\sim0.1$ dex) of the relation above stellar masses of $10^{10}$\,\Msun in transition from \sSFRT (red) to \sSFRD (blue), which is qualitatively in agreement with the results of \citet{Abramson2014}. Below this mass, the two relations are nearly equivalent with a slightly declining slope, as this regime is occupied predominantly by disc-dominated galaxies. 

The flattening of this relation seen at high \Mstar could be explained by the removal of the contribution from (non-star-forming) bulges to the \sSFR which increases with \Mstar \citealt{Abramson2014,Erfanianfar2016}. Although we observe a slight difference in this work when removing the bulge components, it is not sufficient to completely eliminate a residual trend with the total stellar mass. Furthermore, the resulting \sSFRD\;vs.\;\MstarT relation has a comparable scatter to the relation prior to being re-normalised. This is in contrast to the relations shown in \citet{Abramson2014} for which a dramatic increase in $\sigma_{MS}$ is observed. The underlying difference between these analyses is the different structural decomposition catalogues used between these works. \citet{Abramson2014} utilise the \citet{Simard2011} catalogue with fixed \Sersic indices for both the bulge ($n_{Bulge} = 4$) and disc ($n_{Disc} = 1$) components. It has previously been shown that in many cases, this catalogue greatly overestimates the contribution of a bulge component \citep{Meert2015,Cook2019}. In Section \ref{sec:bd_decomp}, we illustrate this in more detail with reference to the catalogue of morphological measurements of \citet{DominguezSanchez2018} determined via a deep-learning algorithm trained upon visually-classified morphological types.

Simply re-normalising the \sSFR by the disc mass does not necessarily identify whether the slope of the \SFRvM relation approaches unity if viewed in the framework of discs alone. This is because in Figure \ref{fig:sSFR_vs_MstarT} above, we are comparing a disc-normalised quantity (\sSFRD) with the \emph{total} stellar mass. For comparison, we also show the equivalent plot for the disc-normalised \sSFR plotted against \MstarD instead of \MstarT (as is the case in e.g. \citealt{Abramson2014}). Here, galaxies do not simply move upwards on the \sSFR axis as in Figure \ref{fig:sSFR_vs_MstarT}, but also towards lower stellar masses. Normalising both axes of this plot by their disc quantities appears to largely remove the flattening of the slope seen in Figure \ref{fig:sSFR_vs_MstarT}. This implies that there is a residual dependence of the \SFMS on stellar mass even for discs taken independently. A slightly declining slope is also observed for galaxies taken as a whole because --- particularly at $M_{\star} < 10^{10}\;M_{\odot}$ --- the majority of their stellar mass resides within a disc component. Indeed, this may explain why a down-bending is observed at high stellar masses with increasing prominence towards later epochs, where galaxies become increasingly dominated by passive bulges \citep{Schreiber2015,Popesso2019a}. Note that the \xGASS sample does not probe enough galaxies in this regime to detect whether the slope is further mass-dependent at these higher stellar masses.

\subsection{Distributions of sSFR across Stellar Mass}
\label{sec:sSFR_distributions}
Much of the difference seen in the slope of the \SFMS between different works likely originates from differences in the definition of ``star-forming'' galaxies. We obtain similar results when reproducing the analysis presented in the previous section with different approaches for defining a ``star-forming'' population (e.g. \NUVr colour; not shown). Note that the definition used in Figure \ref{fig:sSFR_vs_MstarT} is fairly conservative. Placing the division lower might impact the extent of the separation between the \sSFRT and \sSFRD relations as a larger fraction of bulge-dominated galaxies will be included.

We attempt to illustrate whether this could have a significant effect on our results by studying the distributions of \sSFR as a function of stellar mass. The \sSFRvM plane is first divided into $0.4\,dex$ wide bins in stellar mass within which we simultaneously fit Gaussians to both star-forming and passive populations (where present). This is in principle a simplified way of finding the ridge line connecting the peaks of number density along a 2D histogram of galaxies in the \sSFRvM plane as has recently been proposed by \citet{Renzini+Peng2015}. In this way, one obtains the loci of the \SFMS across stellar mass independently of a particular cut in \SFR. Figure \ref{fig:sSFR_hists_by_Mstar} shows the resulting probability density functions obtained when dividing galaxies in bins of \MstarT for both \sSFRT (black) and \sSFRD (blue) in the top row. The distribution of \sSFRD is skewed slightly towards higher values with respect to \sSFRT, confirming that there is indeed an effect when isolating the disc. The peaks of the Gaussian models (vertical dashed lines) trace the \SFMS across stellar mass showing a gradual decline at $M_{\star} \lesssim 10^{10}\;\mathrm{M_{\odot}}$ and a flattening above this. This difference is at the level of 0.1\,dex, similar to that seen in Figure \ref{fig:sSFR_vs_MstarT} and confirms that the presence of a bulge itself cannot entirely explain the negative residual slope in the \sSFRvM relation. Note that the \SFMS becomes highly non-Gaussian towards the high mass end, thus the sharp decline of the \SFMS peak at \MstarT$ \gtrsim 10^{11}\;M_{\odot}$ more likely reflects our inability to fit a suitable Gaussian model to the limited sample size of these data. As a comparison, in the bottom row of Figure \ref{fig:sSFR_hists_by_Mstar}, we replicate this analysis for the $sSFR_{Disc}$ vs $M_{\star,Disc}$ plane as shown in Figure \ref{fig:sSFR_vs_MstarD} by instead binning across disc stellar mass. As before, we confirm that comparing galaxies at a fixed \MstarD removes the differences seen in Figure \ref{fig:sSFR_vs_MstarT}. This shows that the self-similar aspect of discs observed in the previous analysis is not strongly dependent on how star-forming galaxies are selected.

\begin{figure}
    \centering
    \includegraphics[width=\columnwidth]{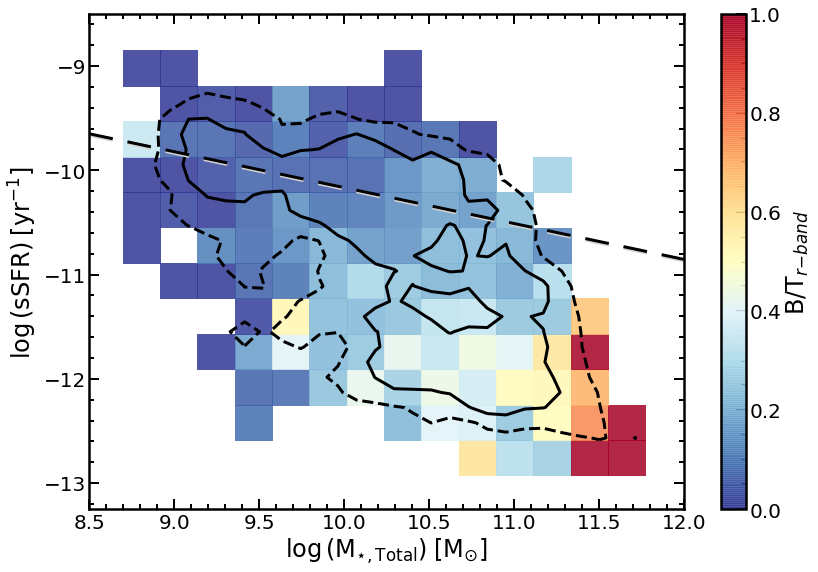}
    \caption{The \sSFRvM plane binned along both axes with the colour showing the mean \rband $B/T$ of galaxies within each bin. The dashed black line corresponds to the \SFMS for the \xGASS sample as defined by \citet{Janowiecki2019}. Contours show the \supth{68} and \supth{95} percentiles containing the data as solid and dashed lines, respectively.}
    \label{fig:sSFR_vs_Mstar_by_B2T}
\end{figure}

\subsection{Role of Bulges Across the Main Sequence}
\label{sec:bulges_across_MS}

\begin{figure*}
    \centering
    \includegraphics[width=0.49\textwidth]{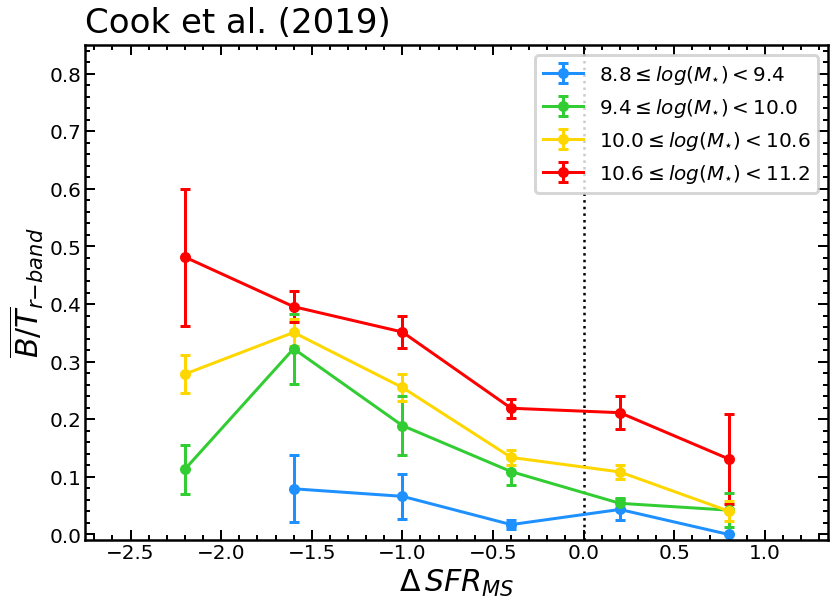}
    \includegraphics[width=0.49\textwidth]{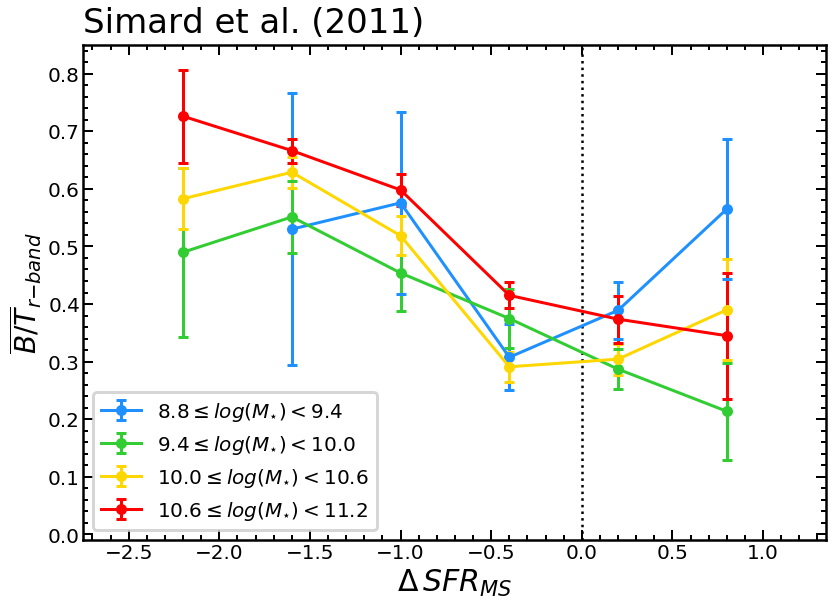}
    \caption{\emph{Left:} Average (\rband) bulge-to-total ratio used in this work plotted against the difference in \sSFR of galaxies from that of the \SFMS ($\Delta SFR$). \emph{Right:} Average (\rband) bulge-to-total ratio measured in \citet{Simard2011} for the same sample plotted against $\Delta SFR$. Coloured lines indicate different ranges in stellar mass with error bars representing the standard errors on the means.}
    \label{fig:B2Tr_vs_DeltaSFMS}
\end{figure*}

A common explanation for the observed slope of the \SFMS is the change in morphology that occurs along it. However, one should not confuse this with the actual passage of individual galaxies within this plane. Rather, the SFRs of galaxies on the \SFMS are regulated in a quasi-steady-state by the inflows and outflows of gas as well as stochastic events such as mergers and violent disc instabilities \citep{Bouche2010,Daddi2010,Genzel2010,Dave2012,Lilly2013,Dekel+Mandelker2014,Tacchella2016}. This has prompted many studies to investigate how various physical properties vary amongst galaxies located above and below the \SFMS, including star formation efficiency, morphology, IR/UV ratio, dust temperature, cold gas content (e.g. \citealt{Wuyts2011,Elbaz2011,Nordon2013,Saintonge2016}). In particular, \citet{Morselli2017,Popesso2019a} observe that the average bulge-to-total ratio of galaxies across the main sequence appears to form a parabolic shape with the minimum locus sitting along the main sequence and the maximum in the passive population. They find intermediate bulge-to-total ratios for galaxies lying slightly below the \SFMS as well as in its upper envelope.

Figure \ref{fig:sSFR_vs_Mstar_by_B2T} shows the \sSFRvM plane with the mean \rband $B/T$ shown in each bin. The overlaid contours represent the $68^\mathrm{th}$ and $95^\mathrm{th}$ percentiles of the number distribution of galaxies. As already pointed out by previous works \citep{Wuyts2011}, this plot shows that the main sequence is primarily populated by disc-dominated galaxies, whereas below the main-sequence there is an increasing contribution from bulges. Along the \SFMS, the mean $B/T$ also increases with stellar mass, varying from $B/T \simeq 0$ at $M_{\star} = 10^{9}$\,\Msun to $B/T \simeq 0.3$ at $M_{\star} = 10^{11}$\,\Msun.

In the left panel of Figure \ref{fig:B2Tr_vs_DeltaSFMS}, we perform a similar exercise as has been done in \citet{Morselli2017} by looking at the mean \rband $B/T$ as a function of distance from the \SFMS ($\Delta SFR_{MS}$). The relatively small size of the \xGASS sample affords separating galaxies into four stellar mass bins of width 0.6\,dex. This is in contrast to the work done by \citet{Morselli2017} who utilise a much larger sample selected from SDSS with the bulge-disc decomposition measurements from \citet{Simard2011}. Here, our results do not replicate the trends with structure as a function of $\Delta SFR_{MS}$, in particular, the average bulge-to-total ratio decreases monotonically from the passive population to the upper envelope of the SFMS. This anticorrelation is observed in each bin of stellar mass, with the exception of the lowest \Mstar bin which remains at a relatively constant average bulge-to-total ratio of $\sim 0.05$.

It is possible that the disparity of the trends observed in Figure \ref{fig:B2Tr_vs_DeltaSFMS} with respect to the \citet{Morselli2017} result may be due to the differences between the samples being studied. Whilst the sample used in \citet{Morselli2017} incorporates $\sim 265,000$ galaxies between $0.02 < z < 0.1$, with $\log\,M_{\star} \geqslant 9.0\;\mathrm{M_{\odot}}$ and identified as not hosting active galactic nuclei, the $\sim1200$ galaxies in \xGASS were selected based on redshift ($0.01 < z < 0.05$) and stellar mass ($9.0 \leqslant \log\,M_{\star} \leqslant 11.5\;\mathrm{M_{\odot}}$) only. The larger sample used in the \citet{Morselli2017} analysis includes rare starburst galaxies with SFRs at more than 1 dex above the \SFMS. Due to the relatively low numbers of galaxies in the \xGASS sample, it is not possible to probe such high SFRs and as such we are not able to make a direct comparison in this regime. However, the upturn in bulge-to-total ratios observed in \citet{Morselli2017} becomes statistically significant for low stellar masses ($\lesssim 10^{10}\;M_{\odot}$) in the regime 0.5 dex above the \SFMS. In \xGASS, $\Delta SFR_{MS} = 0.5$ dex marks the $2\sigma$ confidence interval of star-forming galaxies (i.e. $\sim$95\% of SF galaxies have $\Delta SFR_{MS} \leq 0.5$ dex). That said, we find very few examples of galaxies above this point with $B/T > 0.1$, particular in the lowest stellar masses where the upturn is most evident for \citet{Morselli2017,Popesso2019a}.

To understand how the differences between both samples impacts our results above, we repeat the analysis using the \citet{Simard2011} structural decomposition catalogue with the \xGASS sample in the right panel of Figure \ref{fig:B2Tr_vs_DeltaSFMS}. In all mass bins at all regions across the \SFMS, galaxies show a higher average $B/T$. In at least two of the mass ranges, we observe the trends seen in \citet{Morselli2017} of an upturn in the average $B/T$ of galaxies above $\Delta SFR > 0$. This suggests that, whilst the difference in samples and lower number statistics of xGASS above the main sequence may play a role, the elevated average $B/T$ observed in galaxies above the \SFMS are, at least partially, rooted in the differences in structural decomposition catalogues. We discuss these differences further in Section \ref{sec:bd_decomp}.

\section{Spurious Modelling in Structural Decomposition}
\label{sec:bd_decomp}

\begin{figure}
    \centering
    \includegraphics[width=\columnwidth]{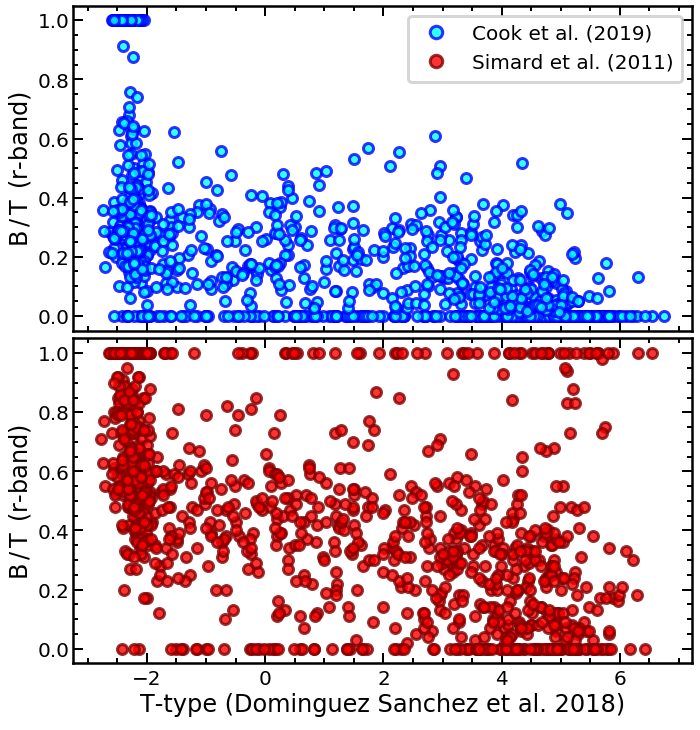}
    \caption{Bulge-to-total ratio measurement derived from the structural decomposition of \xGASS galaxies plotted against their T-types measured in \citet{DominguezSanchez2018}. We compare against structural decomposition catalogues of \citealt{Cook2019} (top) and \citealt{Simard2011} (bottom).}
    \label{fig:B2T_vs_Ttype}
\end{figure}

From the results shown above, it is clear that the structural decomposition of galaxy light profiles and subsequent model classification must be performed very carefully to minimise the contamination of spurious measurements of structural parameters. Relying solely on large number statistics to overcome the many inherent complications of modelling galaxies may prove to be a less viable solution than using smaller samples, where individual galaxies are modelled in greater detail and model selection is based on physical properties rather than statistical measures of their `goodness of fit' (see \citealt{Cook2019}).

\begin{figure}
    \centering
    \includegraphics[width=\columnwidth]{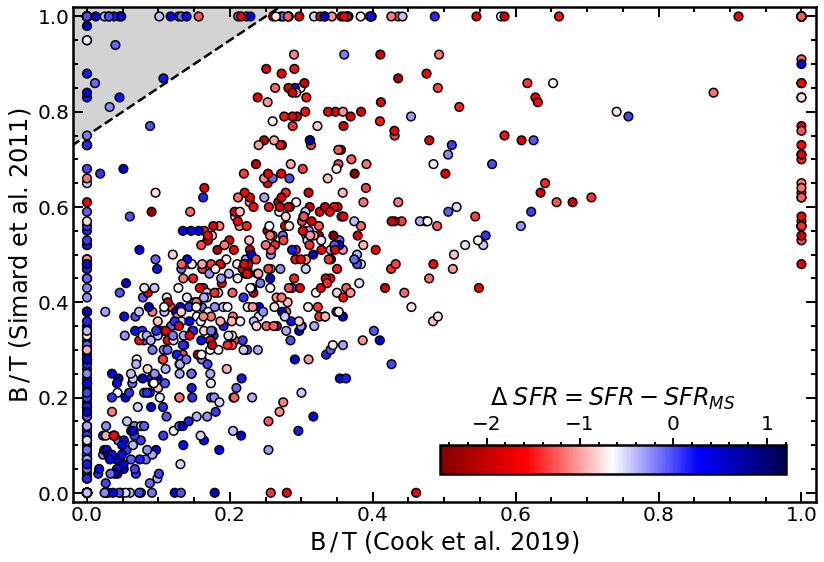}
    \includegraphics[width=0.93\columnwidth]{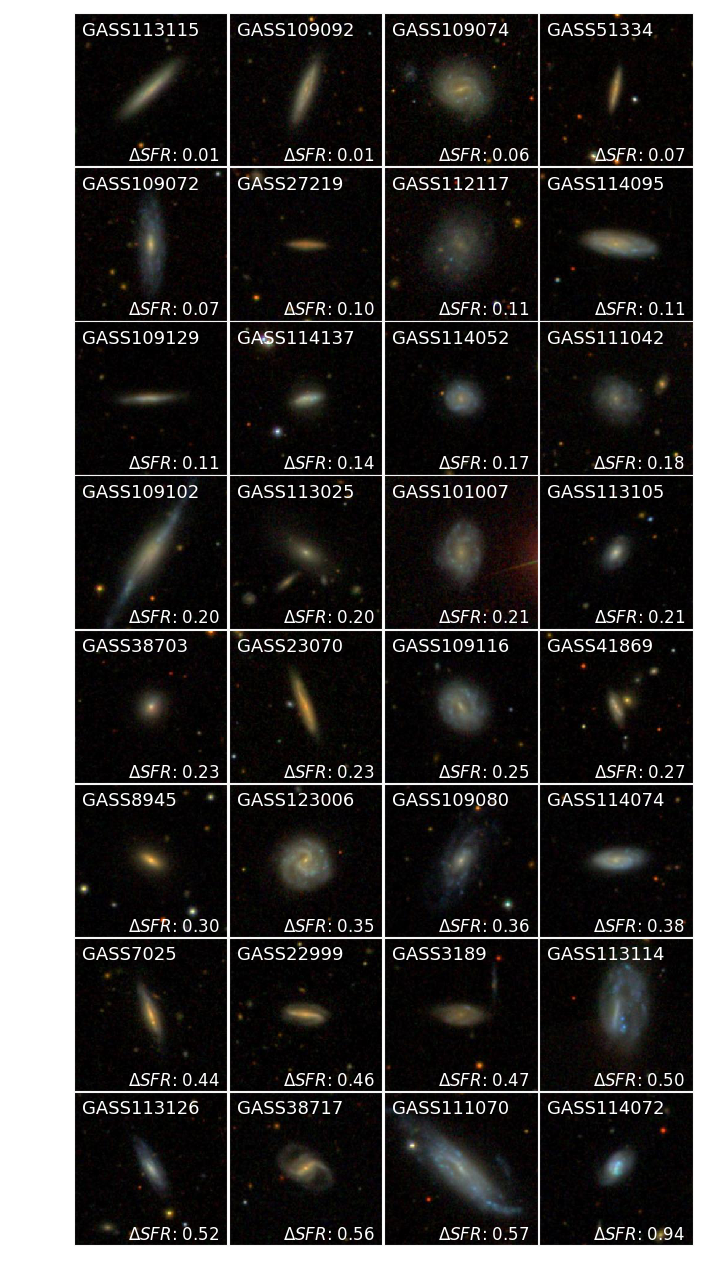}
    \caption{Top panel shows a direct comparison of the \rband bulge-to-total ratios of \xGASS galaxies measured in the \citet{Simard2011} catalogue against those measured in the \citet{Cook2019} catalogue (used in this work). Points are coloured by their distance from the relation defining the star-forming main sequence such that more-positive (bluer) points show galaxies above the \SFMS. The shaded region shows the regime in which the disagreement between the catalogues is greater than 0.75. The panels below show the \SDSS RGB cutout images of galaxies above the \SFMS (i.e. $\Delta SFR > 0$) that fall within this region. With few exceptions, these are predominantly indicative of disc-dominated galaxies.}
    \label{fig:B2T_Cook19_v_Simard11}
\end{figure}

In Figure \ref{fig:B2T_vs_Ttype}, we present a comparison between bulge-to-total ratios derived from the structural decomposition of \SDSS galaxies against their T-types measured from a machine learning algorithm described in \citet{DominguezSanchez2018}. Here, a more-negative T-type indicates an \textit{earlier} morphological class, hence we expect this to correlate with an increasing average $B/T$. This trend is indeed observed to some degree in both catalogues shown here but with important caveats. Firstly, galaxies visually classified as early-type galaxies (T-type $\lesssim -1$) exhibit a large range in $B/T$. This highlights the high degree of uncertainty in modelling light profiles for galaxies that are visually identified as early-type, as profile fitting codes have difficulties distinguishing large spheroids from diffuse discs in this regime. In general, there is a larger spread in $B/T$ at a given T-type in the \citet{Simard2011} catalogue, which again reflects the fact that the sample involved is significantly larger ($\sim 1 \times 10^{6}$ galaxies) and inherently has limitations to the extent to which models can be validated after being modelled. In particular, many galaxies selected from the \citet{Simard2011} catalogue are visually identified as late-type systems from their T-type but have a high B/T. In fact, many of these galaxies would be classified as pure-bulge systems (i.e. $B/T = 1$) when basing the model classification on their \Sersic index ($n > 2.5$) or via the bulge-to-total ratio as measured by a corresponding two-component model. This highlights an important caveat when using structural decomposition catalogues which should be carefully considered before bulge and disc measurements are assigned to a galaxy.

To illustrate this further, Figure \ref{fig:B2T_Cook19_v_Simard11} shows a galaxy-matched comparison between the $B/T$ measured in \citet{Cook2019} and those derived from \citet{Simard2011}. Here, we take the suggested approach from \citet{Simard2011} of using the probability derived from an \textit{F}-test comparing the likelihood that the two-component model is preferred over the single-component one. They assign the value $P_{pS}$ as the \textit{F}-test probability that a two-component model is \emph{not} required compared to that of a pure-\Sersic model. In particular, a galaxy is considered to be best fit by a two-component model if $P_{pS} \leqslant 0.32$. We reiterate here that whilst these single-component models may have provided the mathematically-preferred solution (in terms of their goodness-of-fit metric), given the data quality and model constraints, they may not be physically correct solutions. That said, this probability alone can only indicate whether a particular photometric image exhibits a smaller residual when modelled with one or two components but explicitly does not separate between single-component models that are pure-disc or pure-bulge systems. A cut in \Sersic index is often used as such a discriminator \citep{Allen2006,Meert2015} and, in some cases, an additional measurement of galaxy colour \citep{Kelvin2012}. Here, the cut is made at $n \leq 2.5$ for the \citet{Simard2011} catalogue to remain consistent with previous works and to present a conservative comparison. Whilst the \Sersic index --- or other metrics such as concentration index ($R_{90}/R_{50}$) or central surface brightness --- scales roughly with increasing bulge fraction, the mapping is far more convoluted and certainly not a one-to-one relation \citep{Graham2001b}. In particular, it cannot always distinguish between a purely disc- or spheroid-dominated system, particularly in the case of galaxies that are irregular, disrupted or observed edge-on.

In this particular case, selecting the best model complexity (i.e. number of components) based on residuals of the $data - model$ and \Sersic index is in principle not sufficient. This leads to a non-negligible fraction of inherently pure-disc systems being considered as either bulge-dominated in the case where the \textit{F}-test incorrectly maps to model complexity or as a pure-bulge where the \Sersic index is artificially inflated. The bottom panels of Figure \ref{fig:B2T_Cook19_v_Simard11} show a series of \SDSS RGB cutout images for \xGASS galaxies which show the greatest tension between $B/T$ values derived between the two catalogues. The vast majority of these galaxies are consistent with highly disc-dominated systems corroborating that model selection via a \textit{F}-test probability and cut in \Sersic index is not always valid.

In the era of large surveys, structural decomposition studies will need to move beyond current strategies for model selection in favour for those that can extract more information from the residual ($data - model$) maps. In particular, the use of deep learning algorithms in this field may be the most promising solution to finding the balance needed between a fast, automated and reproducible method that also incorporates the required intuition of visually inspected model selection. In order to implement such deep learning algorithms, one requires a large enough training set that is representative of galaxies in a particular sample. Previous studies have generated mock galaxy images by injecting synthetic \Sersic profiles into real astronomical data. However, these synthetic images do not yet encompass the true complexities of galaxies which vary as a function of environment, mass, projection, sensitivity, redshift, etc. This is becoming increasingly important when investigating the secondary correlations that are present within global galaxy scaling relations.

\section{Discussion}
\label{sec:discussion}
Here, we discuss the main findings of this paper. These are: (a) the declining slope in the \SFMS at high total stellar masses ($M_{\star} > 10^{10}\;M_{\odot}$) is not sufficiently explained by the inclusion of bulges; (b)  importantly, when done in a consistent manner (i.e. \sSFRD vs. \MstarD), this difference is no longer present; (c) the average bulge fraction of galaxies increases monotonically as a function of distance from the \SFMS (at all stellar masses) as well as with increasing total stellar mass.

\begin{figure*}
    \centering
    \includegraphics[width=\textwidth]{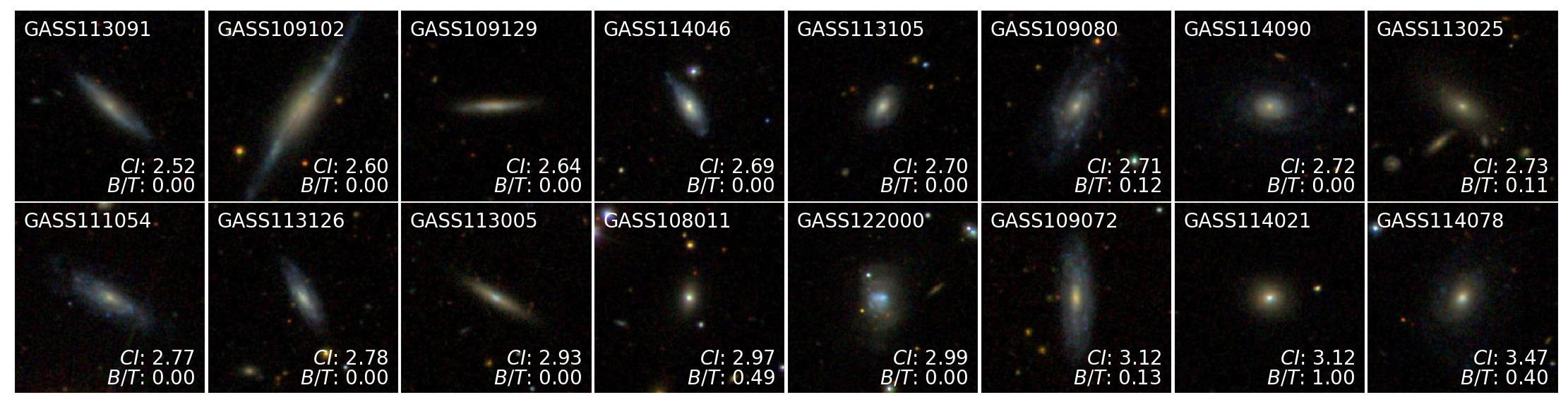}
    \caption{SDSS cutout images centred on low mass ($M_{\star} < 9.4\;M_{\odot}$) \xGASS galaxies above the main sequence with concentration indices ($CI$) of $R_{90}/R_{50} > 2.5$. The (\rband) $B/T$ measured in \citet{Cook2019} and the SDSS-based \rband concentration index for each galaxy are shown in the bottom-right corner. The galaxies are ordered by increasing distance above the \SFMS.}
    \label{fig:high_cindex_cutouts}
\end{figure*}

\subsection{Slope of the Star-forming Main Sequence}
\label{sec:slope_of_SFMS}
Previously, \citet{Abramson2014} have shown that re-normalising specific \SFR by the disc mass alone (as opposed to total stellar mass) can account for $\sim 0.25\,\mathrm{dex}$ of the decline in the \SFMS for every order of magnitude increase in stellar mass for galaxies with $M_{\star} > 10^{10}\;M_{\odot}$. In this work, we have used a different sample and, importantly, a more robust set of structural decomposition measurements to show that the disc-normalised relation (in this same mass regime) accounts for only $\sim 0.1\,\mathrm{dex}$. This result has also been confirmed by \citet{Popesso2019a}, who find that the bulge component accounts for only \perc{10} at $M_{\star} = 10^{10}$ \Msun to \perc{35} at $M_{\star} = 10^{11}$ \Msun. This result alone suggests that the non-identical star formation histories of galaxies encapsulated by the residual negative slope in the \sSFRvM relation cannot be completely explained by the growth of bulges at higher stellar masses.

Moreover, recasting this relation as a disc-normalised quantity (\sSFRD) against a global property (\MstarT) can be difficult to interpret given the residual coupling to $B/T$ that exists between these two quantities. Instead, framing this relation in terms of \MstarD exhibits an identical slope to that seen in the relation with total galaxy quantities. This implies that the observation of higher mass galaxies having built up at earlier epochs and over shorter timescales is itself intrinsic to the disc components of galaxies. Under the assumption that star formation proceeds predominantly within discs, it is perhaps not surprising that the heterogeneous growth of galaxies is rooted in their discs whilst bulges appear to have only at most a secondary role.

In both Figures \ref{fig:sSFR_vs_MstarT} and \ref{fig:sSFR_vs_MstarD}, it is evident that a clear relationship between the \sSFR and stellar mass remains once recast into disc-normalised quantities with a \SFMS scatter comparable to the total relation. This is in contrast to the disc-normalised relations of \citet{Abramson2014} in which the scatter of the \SFMS increases significantly. Although a larger sample was used, the greater uncertainties in the bulge and disc model parameters leads to a larger overall scatter and the resulting relation between $SFR/M_{\star,Disc}$ and \MstarT is no longer well-defined. We have shown that with a smaller sample, the resulting scatter can be significantly reduced when using structural decomposition measurements with further model validation and filtering as described in \citet{Cook2019}.

\subsection{Morphology across the SFMS}
\label{sec:morph_along_MS}
Perpendicular to investigating the role of bulges in regulating the relation along the \SFMS, we also investigate the distribution of morphology across it. The trends of an increasing bulge prominence as a function of distance below the \SFMS observed in this work are in good agreement with recent studies \citep{Bluck2014,Morselli2019}. At all stellar masses, galaxies below the main sequence are always on average more bulge-dominated than those on or above it. Whilst at face value this might suggest that the build up of a bulge may be linked to the suppression of star formation in galaxies, \citet{Cook2019} showed that the presence (and relative prominence) of bulges in star-forming galaxies has little-to-no impact on their overall atomic hydrogen (\HI) gas reservoirs. If we consider the \HI reservoir within a galaxy to indicate its potential for future star formation, we would expect a signature to appear here if, in fact, the presence of a bulge could in some way affect the continued star formation in a galaxy.

Note that our results are in tension with those found in previous studies (e.g. \citealt{Morselli2017,Popesso2019a}) also using local samples of \SDSS galaxies. In particular, \citet{Morselli2017} find that at most stellar masses, the minimum average $B/T$ of galaxies aligns closely with the peak of the \SFMS (i.e. $\Delta SFR_{MS} \equiv 0$), whereas \citet{Popesso2019a} find this minimum to increase gradually as a function of stellar mass. Note, however, that using global structural parameters of CANDELS galaxies modelled by \citet{vanderWel2012}, \citet{Morselli2019} showed recently that galaxies above the \SFMS exhibit lower \Sersic indices on average than galaxies on or below the \SFMS, which is consistent with this work. Our results in Figure \ref{fig:B2Tr_vs_DeltaSFMS} show that the average $B/T$ is monotonically declining with no apparent increase above the \SFMS. This implies that as galaxies experience episodes of heightened star formation activity, a change in their morphology does not necessarily follow. The results of \citet{Morselli2017,Popesso2019a} have prompted a scenario in which galaxies receive a large infall of gas that triggers star formation and subsequently promotes the growth of the central bulge. In this compaction-depletion scenario, repeated phases of gas inflow towards the centre of a galaxy are followed by depletion due to an episode of heightened star formation activity. This advances galaxies along the main sequence before reaching a final quenching event. Our data instead do not reveal a population of starburst galaxies (emerging from either merger events or violent disc instabilities) which harbour a prominent star-forming bulge component.

As is highlighted by Figure \ref{fig:B2T_Cook19_v_Simard11}, this tension is likely due to the differences between structural decomposition measurements used in these studies, rather than differences in calibration of \SFR indicators or stellar mass measurements. The presence of visually classified late-type galaxies modelled as highly bulge-dominated systems partly explains the discrepancies between this study and those based on the \citet{Simard2011} catalogue.

Lastly, we note that a similar increase above the main sequence has been observed when using SDSS-based concentration indices ($R_{90}/R_{50}$) in place of of structural decomposition (e.g. Appendix B of \citealt{Morselli2017}). It is important to emphasise that from a physical point of view there is no one-to-one mapping between concentration index and $B/T$ and even the mapping onto \Sersic index is not strictly monotonic when the effects of seeing are considered \citep{Graham+Driver2005}. Moreover, the concentration index (and accordingly, a single-component \Sersic index) is itself a poor proxy for $B/T$ since both low and high $R_{90}/R_{50}$ map to single-component models as pure-discs and pure-bulges, respectively. As an example, low mass star-forming galaxies often show features such as inner star-bursts or bars which increase the concentration of light (and \Sersic index) measured despite there being no bulge structure present. This is highlighted in Figure \ref{fig:high_cindex_cutouts}, where we show examples of low mass galaxies in \xGASS above the main sequence and with concentration indices greater than 2.5 which roughly delineates where the upturn becomes statistically significant in the \citet{Morselli2017} analysis. As can be seen, our structural decomposition suggests that in most cases, there is little-to-no contribution from a bulge structure but a large concentration index results nonetheless. As such, it should not be used as evidence that highly star-forming systems have bigger bulges.

\section{Conclusions}
\label{sec:conclusion}
In this work, we study the role of bulges both as regulators of star formation rate (along the \SFMS) as well as by-products (across the \SFMS) using our catalogue of robust structural decompositions \citep{Cook2019} of the \xGASS sample. We find that the slope of the \sSFRvM flattens by 0.1\,dex per decade in \Mstar when re-normalising the specific star formation rate by the stellar mass of the disc. However, this flattening is only observed above a stellar mass of $M_{\star} \sim 10^{10}\;M_{\odot}$; below this, the relation retains a gradual negative slope as this regime is dominated by pure-disc systems. This fact, in addition to the persistence of the negative slope when plotting \sSFRD against the disc stellar mass, indicates that the residual mass dependence of this relation is more closely linked to physical processes acting on the disc, rather than the contribution from bulges. Galaxies situated on the \SFMS exhibit an increasing average $B/T$ as a function of stellar mass, from nearly pure-discs ($B/T \sim 0$) to \simperc{30} bulge fractions over the mass range of $M_{\star} = 10^{9}$\,--\,$10^{11.25}\;M_{\odot}$. However, this alone is not sufficient to explain the residual slope in the \sSFRvM relation.

Furthermore, we have found that the average $B/T$ of galaxies as a function of distance from the \SFMS is monotonically decreasing at all stellar masses; with the exception of our lowest bin ($M_{\star} < 10^{9.4}\;M_{\odot}$) which is consistent with pure-disc systems throughout its range of star-formation rates. This finding is in agreement with some previous works \citep{Bluck2014,Morselli2019} but not with other works which find an increased average $B/T$ above the \SFMS \citep{Morselli2017,Popesso2019a}. We do not find evidence for a population of starburst galaxies with systematically higher bulge fractions in the local Universe.

We attribute this discrepancy to differences in the structural decomposition measurements used in each of these works. The limited model validation viable in such large catalogues can lead to spurious structural measurements being assigned to galaxies. In particular, using a goodness-of-fit metric to decide model complexity (number of components) and a proxy for morphology (e.g. \Sersic index, concentration index) to distinguish pure-discs and pure-bulges is not in itself sufficient to classify galaxies in a robust manner.

\section*{Acknowledgements}
We would like to gratefully thank the referee for their insightful and constructive comments.

Parts of this research were supported by the Australian Research Council Centre of Excellence for All Sky Astrophysics in 3 Dimensions (ASTRO 3D), through project number CE170100013.

LC is the recipient of an Australian Research Council Future Fellowship (FT180100066) funded by the Australian Government.

\bibliographystyle{mnras.bst}
\bibliography{bibfile.bib}

\begin{thebibliography}{}
\makeatletter
\relax
\def\mn@urlcharsother{\let\do\@makeother \do\$\do\&\do\#\do\^\do\_\do\%\do\~}
\def\mn@doi{\begingroup\mn@urlcharsother \@ifnextchar [ {\mn@doi@}
  {\mn@doi@[]}}
\def\mn@doi@[#1]#2{\def\@tempa{#1}\ifx\@tempa\@empty \href
  {http://dx.doi.org/#2} {doi:#2}\else \href {http://dx.doi.org/#2} {#1}\fi
  \endgroup}
\def\mn@eprint#1#2{\mn@eprint@#1:#2::\@nil}
\def\mn@eprint@arXiv#1{\href {http://arxiv.org/abs/#1} {{\tt arXiv:#1}}}
\def\mn@eprint@dblp#1{\href {http://dblp.uni-trier.de/rec/bibtex/#1.xml}
  {dblp:#1}}
\def\mn@eprint@#1:#2:#3:#4\@nil{\def\@tempa {#1}\def\@tempb {#2}\def\@tempc
  {#3}\ifx \@tempc \@empty \let \@tempc \@tempb \let \@tempb \@tempa \fi \ifx
  \@tempb \@empty \def\@tempb {arXiv}\fi \@ifundefined
  {mn@eprint@\@tempb}{\@tempb:\@tempc}{\expandafter \expandafter \csname
  mn@eprint@\@tempb\endcsname \expandafter{\@tempc}}}

\bibitem[\protect\citeauthoryear{{Abazajian} et~al.,}{{Abazajian}
  et~al.}{2009}]{Abazajian2009}
{Abazajian} K.~N.,  et~al., 2009, \mn@doi [The Astrophysical Journal Supplement
  Series] {10.1088/0067-0049/182/2/543}, \href
  {https://ui.adsabs.harvard.edu/\#abs/2009ApJS..182..543A} {182, 543}

\bibitem[\protect\citeauthoryear{{Abramson}, {Kelson}, {Dressler}, {Poggianti},
  {Gladders}, {Oemler}  \& {Vulcani}}{{Abramson} et~al.}{2014}]{Abramson2014}
{Abramson} L.~E.,  {Kelson} D.~D.,  {Dressler} A.,  {Poggianti} B.,  {Gladders}
  M.~D.,  {Oemler} Augustus J.,   {Vulcani} B.,  2014, \mn@doi [\apj]
  {10.1088/2041-8205/785/2/L36}, \href
  {https://ui.adsabs.harvard.edu/\#abs/2014ApJ...785L..36A} {785, L36}

\bibitem[\protect\citeauthoryear{{Adelman-McCarthy} et~al.,}{{Adelman-McCarthy}
  et~al.}{2008}]{Adelman-McCarthy2008}
{Adelman-McCarthy} J.~K.,  et~al., 2008, \mn@doi [The Astrophysical Journal
  Supplement Series] {10.1086/524984}, \href
  {https://ui.adsabs.harvard.edu/\#abs/2008ApJS..175..297A} {175, 297}

\bibitem[\protect\citeauthoryear{{Allen}, {Driver}, {Graham}, {Cameron},
  {Liske}  \& {de Propris}}{{Allen} et~al.}{2006}]{Allen2006}
{Allen} P.~D.,  {Driver} S.~P.,  {Graham} A.~W.,  {Cameron} E.,  {Liske} J.,
  {de Propris} R.,  2006, \mn@doi [\mnras] {10.1111/j.1365-2966.2006.10586.x},
  \href {http://adsabs.harvard.edu/abs/2006MNRAS.371....2A} {371, 2}

\bibitem[\protect\citeauthoryear{{Belfiore} et~al.,}{{Belfiore}
  et~al.}{2018}]{Belfiore2018}
{Belfiore} F.,  et~al., 2018, \mn@doi [\mnras] {10.1093/mnras/sty768}, \href
  {https://ui.adsabs.harvard.edu/abs/2018MNRAS.477.3014B} {477, 3014}

\bibitem[\protect\citeauthoryear{{Bluck}, {Mendel}, {Ellison}, {Moreno},
  {Simard}, {Patton}  \& {Starkenburg}}{{Bluck} et~al.}{2014}]{Bluck2014}
{Bluck} A. F.~L.,  {Mendel} J.~T.,  {Ellison} S.~L.,  {Moreno} J.,  {Simard}
  L.,  {Patton} D.~R.,   {Starkenburg} E.,  2014, \mn@doi [\mnras]
  {10.1093/mnras/stu594}, \href
  {https://ui.adsabs.harvard.edu/abs/2014MNRAS.441..599B} {441, 599}

\bibitem[\protect\citeauthoryear{{Boselli}, {Gavazzi}, {Donas}  \&
  {Scodeggio}}{{Boselli} et~al.}{2001}]{Boselli2001}
{Boselli} A.,  {Gavazzi} G.,  {Donas} J.,   {Scodeggio} M.,  2001, \mn@doi
  [\aj] {10.1086/318734}, \href
  {https://ui.adsabs.harvard.edu/abs/2001AJ....121..753B} {121, 753}

\bibitem[\protect\citeauthoryear{{Bouch{\'e}} et~al.,}{{Bouch{\'e}}
  et~al.}{2010}]{Bouche2010}
{Bouch{\'e}} N.,  et~al., 2010, \mn@doi [\apj] {10.1088/0004-637X/718/2/1001},
  \href {https://ui.adsabs.harvard.edu/abs/2010ApJ...718.1001B} {718, 1001}

\bibitem[\protect\citeauthoryear{{Brinchmann}, {Charlot}, {White}, {Tremonti},
  {Kauffmann}, {Heckman}  \& {Brinkmann}}{{Brinchmann}
  et~al.}{2004}]{Brinchmann2004}
{Brinchmann} J.,  {Charlot} S.,  {White} S.~D.~M.,  {Tremonti} C.,  {Kauffmann}
  G.,  {Heckman} T.,   {Brinkmann} J.,  2004, \mn@doi [\mnras]
  {10.1111/j.1365-2966.2004.07881.x}, \href
  {https://ui.adsabs.harvard.edu/#abs/2004MNRAS.351.1151B} {351, 1151}

\bibitem[\protect\citeauthoryear{{Catinella} et~al.,}{{Catinella}
  et~al.}{2010}]{Catinella2010}
{Catinella} B.,  et~al., 2010, \mn@doi [\mnras]
  {10.1111/j.1365-2966.2009.16180.x}, \href
  {http://adsabs.harvard.edu/abs/2010MNRAS.403..683C} {403, 683}

\bibitem[\protect\citeauthoryear{{Catinella} et~al.,}{{Catinella}
  et~al.}{2018}]{Catinella2018}
{Catinella} B.,  et~al., 2018, \mn@doi [\mnras] {10.1093/mnras/sty089}, \href
  {https://ui.adsabs.harvard.edu/\#abs/2018MNRAS.476..875C} {476, 875}

\bibitem[\protect\citeauthoryear{{Cook}, {Cortese}, {Catinella}  \&
  {Robotham}}{{Cook} et~al.}{2019}]{Cook2019}
{Cook} R. H.~W.,  {Cortese} L.,  {Catinella} B.,   {Robotham} A. S.~G.,  2019,
  arXiv e-prints, \href {https://ui.adsabs.harvard.edu/abs/2019arXiv190910202C}
  {p. arXiv:1909.10202}

\bibitem[\protect\citeauthoryear{{Daddi} et~al.,}{{Daddi}
  et~al.}{2007}]{Daddi2007}
{Daddi} E.,  et~al., 2007, \mn@doi [\apj] {10.1086/521818}, \href
  {https://ui.adsabs.harvard.edu/#abs/2007ApJ...670..156D} {670, 156}

\bibitem[\protect\citeauthoryear{{Daddi} et~al.,}{{Daddi}
  et~al.}{2010}]{Daddi2010}
{Daddi} E.,  et~al., 2010, \mn@doi [\apjl] {10.1088/2041-8205/714/1/L118},
  \href {https://ui.adsabs.harvard.edu/abs/2010ApJ...714L.118D} {714, L118}

\bibitem[\protect\citeauthoryear{{Dav{\'e}}, {Finlator}  \&
  {Oppenheimer}}{{Dav{\'e}} et~al.}{2012}]{Dave2012}
{Dav{\'e}} R.,  {Finlator} K.,   {Oppenheimer} B.~D.,  2012, \mn@doi [\mnras]
  {10.1111/j.1365-2966.2011.20148.x}, \href
  {https://ui.adsabs.harvard.edu/abs/2012MNRAS.421...98D} {421, 98}

\bibitem[\protect\citeauthoryear{{Davies} et~al.,}{{Davies}
  et~al.}{2016}]{Davies2016}
{Davies} L.~J.~M.,  et~al., 2016, \mn@doi [\mnras] {10.1093/mnras/stw1342},
  \href {https://ui.adsabs.harvard.edu/abs/2016MNRAS.461..458D} {461, 458}

\bibitem[\protect\citeauthoryear{{Dekel} \& {Mandelker}}{{Dekel} \&
  {Mandelker}}{2014}]{Dekel+Mandelker2014}
{Dekel} A.,  {Mandelker} N.,  2014, \mn@doi [\mnras] {10.1093/mnras/stu1427},
  \href {https://ui.adsabs.harvard.edu/abs/2014MNRAS.444.2071D} {444, 2071}

\bibitem[\protect\citeauthoryear{{Dom{\'\i}nguez S{\'a}nchez},
  {Huertas-Company}, {Bernardi}, {Tuccillo}  \& {Fischer}}{{Dom{\'\i}nguez
  S{\'a}nchez} et~al.}{2018}]{DominguezSanchez2018}
{Dom{\'\i}nguez S{\'a}nchez} H.,  {Huertas-Company} M.,  {Bernardi} M.,
  {Tuccillo} D.,   {Fischer} J.~L.,  2018, \mn@doi [\mnras]
  {10.1093/mnras/sty338}, \href
  {https://ui.adsabs.harvard.edu/abs/2018MNRAS.476.3661D} {476, 3661}

\bibitem[\protect\citeauthoryear{{Elbaz} et~al.,}{{Elbaz}
  et~al.}{2007}]{Elbaz2007}
{Elbaz} D.,  et~al., 2007, \mn@doi [\aap] {10.1051/0004-6361:20077525}, \href
  {https://ui.adsabs.harvard.edu/\#abs/2007A&A...468...33E} {468, 33}

\bibitem[\protect\citeauthoryear{{Elbaz} et~al.,}{{Elbaz}
  et~al.}{2011}]{Elbaz2011}
{Elbaz} D.,  et~al., 2011, \mn@doi [\aap] {10.1051/0004-6361/201117239}, \href
  {https://ui.adsabs.harvard.edu/\#abs/2011A&A...533A.119E} {533, A119}

\bibitem[\protect\citeauthoryear{{Ellison}, {S{\'a}nchez}, {Ibarra-Medel},
  {Antonio}, {Mendel}  \& {Barrera-Ballesteros}}{{Ellison}
  et~al.}{2018}]{Ellison2018a}
{Ellison} S.~L.,  {S{\'a}nchez} S.~F.,  {Ibarra-Medel} H.,  {Antonio} B.,
  {Mendel} J.~T.,   {Barrera-Ballesteros} J.,  2018, \mn@doi [\mnras]
  {10.1093/mnras/stx2882}, \href
  {https://ui.adsabs.harvard.edu/abs/2018MNRAS.474.2039E} {474, 2039}

\bibitem[\protect\citeauthoryear{{Erfanianfar} et~al.,}{{Erfanianfar}
  et~al.}{2016}]{Erfanianfar2016}
{Erfanianfar} G.,  et~al., 2016, \mn@doi [\mnras] {10.1093/mnras/stv2485},
  \href {https://ui.adsabs.harvard.edu/abs/2016MNRAS.455.2839E} {455, 2839}

\bibitem[\protect\citeauthoryear{{Gavazzi} et~al.,}{{Gavazzi}
  et~al.}{2015}]{Gavazzi2015}
{Gavazzi} G.,  et~al., 2015, \mn@doi [\aap] {10.1051/0004-6361/201425351},
  \href {https://ui.adsabs.harvard.edu/abs/2015A&A...580A.116G} {580, A116}

\bibitem[\protect\citeauthoryear{{Genzel} et~al.,}{{Genzel}
  et~al.}{2010}]{Genzel2010}
{Genzel} R.,  et~al., 2010, \mn@doi [\mnras]
  {10.1111/j.1365-2966.2010.16969.x}, \href
  {https://ui.adsabs.harvard.edu/abs/2010MNRAS.407.2091G} {407, 2091}

\bibitem[\protect\citeauthoryear{{Graham} \& {Driver}}{{Graham} \&
  {Driver}}{2005}]{Graham+Driver2005}
{Graham} A.~W.,  {Driver} S.~P.,  2005, \mn@doi [Publications of the
  Astronomical Society of Australia] {10.1071/AS05001}, \href
  {https://ui.adsabs.harvard.edu/#abs/2005PASA...22..118G} {22, 118}

\bibitem[\protect\citeauthoryear{{Graham}, {Trujillo}  \& {Caon}}{{Graham}
  et~al.}{2001}]{Graham2001b}
{Graham} A.~W.,  {Trujillo} I.,   {Caon} N.,  2001, \mn@doi [\aj]
  {10.1086/323090}, \href
  {https://ui.adsabs.harvard.edu/abs/2001AJ....122.1707G} {122, 1707}

\bibitem[\protect\citeauthoryear{{Guo}, {Zheng}, {Wang}  \& {Fu}}{{Guo}
  et~al.}{2015}]{Guo2015}
{Guo} K.,  {Zheng} X.~Z.,  {Wang} T.,   {Fu} H.,  2015, \mn@doi [\apj]
  {10.1088/2041-8205/808/2/L49}, \href
  {https://ui.adsabs.harvard.edu/\#abs/2015ApJ...808L..49G} {808, L49}

\bibitem[\protect\citeauthoryear{{Hopkins} \& {Beacom}}{{Hopkins} \&
  {Beacom}}{2006}]{Hopkins2006b}
{Hopkins} A.~M.,  {Beacom} J.~F.,  2006, \mn@doi [\apj] {10.1086/506610}, \href
  {https://ui.adsabs.harvard.edu/abs/2006ApJ...651..142H} {651, 142}

\bibitem[\protect\citeauthoryear{{Janowiecki}, {Catinella}, {Cortese},
  {Saintonge}, {Brown}  \& {Wang}}{{Janowiecki} et~al.}{2017}]{Janowiecki2017}
{Janowiecki} S.,  {Catinella} B.,  {Cortese} L.,  {Saintonge} A.,  {Brown} T.,
   {Wang} J.,  2017, \mn@doi [\mnras] {10.1093/mnras/stx046}, \href
  {https://ui.adsabs.harvard.edu/\#abs/2017MNRAS.466.4795J} {466, 4795}

\bibitem[\protect\citeauthoryear{{Janowiecki}, {Catinella}  \&
  {Cortese}}{{Janowiecki} et~al.}{2019}]{Janowiecki2019}
{Janowiecki} S.,  {Catinella} B.,   {Cortese} L.,  2019, in American
  Astronomical Society Meeting Abstracts \#233. p. 429.04

\bibitem[\protect\citeauthoryear{{Kelvin} et~al.,}{{Kelvin}
  et~al.}{2012}]{Kelvin2012}
{Kelvin} L.~S.,  et~al., 2012, \mn@doi [\mnras]
  {10.1111/j.1365-2966.2012.20355.x}, \href
  {https://ui.adsabs.harvard.edu/\#abs/2012MNRAS.421.1007K} {421, 1007}

\bibitem[\protect\citeauthoryear{{Kennicutt}, {Tamblyn}  \&
  {Congdon}}{{Kennicutt} et~al.}{1994}]{Kennicutt1994}
{Kennicutt} Robert~C. J.,  {Tamblyn} P.,   {Congdon} C.~E.,  1994, \mn@doi
  [\apj] {10.1086/174790}, \href
  {https://ui.adsabs.harvard.edu/abs/1994ApJ...435...22K} {435, 22}

\bibitem[\protect\citeauthoryear{{Lee} et~al.,}{{Lee} et~al.}{2015}]{Lee2015}
{Lee} N.,  et~al., 2015, \mn@doi [\apj] {10.1088/0004-637X/801/2/80}, \href
  {https://ui.adsabs.harvard.edu/\#abs/2015ApJ...801...80L} {801, 80}

\bibitem[\protect\citeauthoryear{{Lilly}, {Le Fevre}, {Hammer}  \&
  {Crampton}}{{Lilly} et~al.}{1996}]{Lilly1996}
{Lilly} S.~J.,  {Le Fevre} O.,  {Hammer} F.,   {Crampton} D.,  1996, \mn@doi
  [\apjl] {10.1086/309975}, \href
  {https://ui.adsabs.harvard.edu/abs/1996ApJ...460L...1L} {460, L1}

\bibitem[\protect\citeauthoryear{{Lilly}, {Carollo}, {Pipino}, {Renzini}  \&
  {Peng}}{{Lilly} et~al.}{2013}]{Lilly2013}
{Lilly} S.~J.,  {Carollo} C.~M.,  {Pipino} A.,  {Renzini} A.,   {Peng} Y.,
  2013, \mn@doi [\apj] {10.1088/0004-637X/772/2/119}, \href
  {https://ui.adsabs.harvard.edu/abs/2013ApJ...772..119L} {772, 119}

\bibitem[\protect\citeauthoryear{{Madau}, {Pozzetti}  \& {Dickinson}}{{Madau}
  et~al.}{1998}]{Madau1998}
{Madau} P.,  {Pozzetti} L.,   {Dickinson} M.,  1998, \mn@doi [\apj]
  {10.1086/305523}, \href
  {https://ui.adsabs.harvard.edu/abs/1998ApJ...498..106M} {498, 106}

\bibitem[\protect\citeauthoryear{{Martin} et~al.,}{{Martin}
  et~al.}{2005}]{Martin2005}
{Martin} D.~C.,  et~al., 2005, \mn@doi [\apj] {10.1086/426387}, \href
  {https://ui.adsabs.harvard.edu/#abs/2005ApJ...619L...1M} {619, L1}

\bibitem[\protect\citeauthoryear{{Meert}, {Vikram}  \& {Bernardi}}{{Meert}
  et~al.}{2015}]{Meert2015}
{Meert} A.,  {Vikram} V.,   {Bernardi} M.,  2015, \mn@doi [\mnras]
  {10.1093/mnras/stu2333}, \href
  {https://ui.adsabs.harvard.edu/#abs/2015MNRAS.446.3943M} {446, 3943}

\bibitem[\protect\citeauthoryear{{Morselli}, {Popesso}, {Erfanianfar}  \&
  {Concas}}{{Morselli} et~al.}{2017}]{Morselli2017}
{Morselli} L.,  {Popesso} P.,  {Erfanianfar} G.,   {Concas} A.,  2017, \mn@doi
  [\aap] {10.1051/0004-6361/201629409}, \href
  {https://ui.adsabs.harvard.edu/abs/2017A&A...597A..97M} {597, A97}

\bibitem[\protect\citeauthoryear{{Morselli}, {Popesso}, {Cibinel}, {Oesch},
  {Montes}, {Atek}, {Illingworth}  \& {Holden}}{{Morselli}
  et~al.}{2019}]{Morselli2019}
{Morselli} L.,  {Popesso} P.,  {Cibinel} A.,  {Oesch} P.~A.,  {Montes} M.,
  {Atek} H.,  {Illingworth} G.~D.,   {Holden} B.,  2019, \mn@doi [\aap]
  {10.1051/0004-6361/201834559}, \href
  {https://ui.adsabs.harvard.edu/abs/2019A&A...626A..61M} {626, A61}

\bibitem[\protect\citeauthoryear{{Neistein}, {van den Bosch}  \&
  {Dekel}}{{Neistein} et~al.}{2006}]{Neistein2006}
{Neistein} E.,  {van den Bosch} F.~C.,   {Dekel} A.,  2006, \mn@doi [\mnras]
  {10.1111/j.1365-2966.2006.10918.x}, \href
  {https://ui.adsabs.harvard.edu/abs/2006MNRAS.372..933N} {372, 933}

\bibitem[\protect\citeauthoryear{{Noeske} et~al.,}{{Noeske}
  et~al.}{2007}]{Noeske2007}
{Noeske} K.~G.,  et~al., 2007, \mn@doi [\apj] {10.1086/517926}, \href
  {https://ui.adsabs.harvard.edu/\#abs/2007ApJ...660L..43N} {660, L43}

\bibitem[\protect\citeauthoryear{{Nordon} et~al.,}{{Nordon}
  et~al.}{2013}]{Nordon2013}
{Nordon} R.,  et~al., 2013, \mn@doi [\apj] {10.1088/0004-637X/762/2/125}, \href
  {https://ui.adsabs.harvard.edu/abs/2013ApJ...762..125N} {762, 125}

\bibitem[\protect\citeauthoryear{{Pannella} et~al.,}{{Pannella}
  et~al.}{2009}]{Pannella2009}
{Pannella} M.,  et~al., 2009, \mn@doi [\apjl] {10.1088/0004-637X/698/2/L116},
  \href {https://ui.adsabs.harvard.edu/abs/2009ApJ...698L.116P} {698, L116}

\bibitem[\protect\citeauthoryear{{Peng} et~al.,}{{Peng}
  et~al.}{2010}]{Peng2010b}
{Peng} Y.-j.,  et~al., 2010, \mn@doi [\apj] {10.1088/0004-637X/721/1/193},
  \href {https://ui.adsabs.harvard.edu/abs/2010ApJ...721..193P} {721, 193}

\bibitem[\protect\citeauthoryear{{Popesso} et~al.,}{{Popesso}
  et~al.}{2019a}]{Popesso2019a}
{Popesso} P.,  et~al., 2019a, \mn@doi [Monthly Notices of the Royal
  Astronomical Society] {10.1093/mnras/sty3210}, \href
  {https://ui.adsabs.harvard.edu/abs/2019MNRAS.483.3213P} {483, 3213}

\bibitem[\protect\citeauthoryear{{Popesso} et~al.,}{{Popesso}
  et~al.}{2019b}]{Popesso2019b}
{Popesso} P.,  et~al., 2019b, \mn@doi [\mnras] {10.1093/mnras/stz2635}, \href
  {https://ui.adsabs.harvard.edu/abs/2019MNRAS.490.5285P} {490, 5285}

\bibitem[\protect\citeauthoryear{{Renzini} \& {Peng}}{{Renzini} \&
  {Peng}}{2015}]{Renzini+Peng2015}
{Renzini} A.,  {Peng} Y.-j.,  2015, \mn@doi [\apjl]
  {10.1088/2041-8205/801/2/L29}, \href
  {https://ui.adsabs.harvard.edu/abs/2015ApJ...801L..29R} {801, L29}

\bibitem[\protect\citeauthoryear{{Robotham}, {Taranu}, {Tobar}, {Moffett}  \&
  {Driver}}{{Robotham} et~al.}{2017}]{Robotham2017}
{Robotham} A.~S.~G.,  {Taranu} D.~S.,  {Tobar} R.,  {Moffett} A.,   {Driver}
  S.~P.,  2017, \mn@doi [\mnras] {10.1093/mnras/stw3039}, \href
  {http://adsabs.harvard.edu/abs/2017MNRAS.466.1513R} {466, 1513}

\bibitem[\protect\citeauthoryear{{Rodighiero} et~al.,}{{Rodighiero}
  et~al.}{2010}]{Rodighiero2010}
{Rodighiero} G.,  et~al., 2010, \mn@doi [\aap] {10.1051/0004-6361/201014624},
  \href {https://ui.adsabs.harvard.edu/\#abs/2010A&A...518L..25R} {518, L25}

\bibitem[\protect\citeauthoryear{{Rodighiero} et~al.,}{{Rodighiero}
  et~al.}{2014}]{Rodighiero2014}
{Rodighiero} G.,  et~al., 2014, \mn@doi [\mnras] {10.1093/mnras/stu1110}, \href
  {https://ui.adsabs.harvard.edu/abs/2014MNRAS.443...19R} {443, 19}

\bibitem[\protect\citeauthoryear{{Saintonge} et~al.,}{{Saintonge}
  et~al.}{2016}]{Saintonge2016}
{Saintonge} A.,  et~al., 2016, \mn@doi [\mnras] {10.1093/mnras/stw1715}, \href
  {https://ui.adsabs.harvard.edu/abs/2016MNRAS.462.1749S} {462, 1749}

\bibitem[\protect\citeauthoryear{{Salim} et~al.,}{{Salim}
  et~al.}{2007}]{Salim2007}
{Salim} S.,  et~al., 2007, \mn@doi [The Astrophysical Journal Supplement
  Series] {10.1086/519218}, \href
  {https://ui.adsabs.harvard.edu/\#abs/2007ApJS..173..267S} {173, 267}

\bibitem[\protect\citeauthoryear{{Schreiber} et~al.,}{{Schreiber}
  et~al.}{2015}]{Schreiber2015}
{Schreiber} C.,  et~al., 2015, \mn@doi [\aap] {10.1051/0004-6361/201425017},
  \href {https://ui.adsabs.harvard.edu/abs/2015A&A...575A..74S} {575, A74}

\bibitem[\protect\citeauthoryear{{Simard}, {Mendel}, {Patton}, {Ellison}  \&
  {McConnachie}}{{Simard} et~al.}{2011}]{Simard2011}
{Simard} L.,  {Mendel} J.~T.,  {Patton} D.~R.,  {Ellison} S.~L.,
  {McConnachie} A.~W.,  2011, \mn@doi [The Astrophysical Journal Supplement
  Series] {10.1088/0067-0049/196/1/11}, \href
  {https://ui.adsabs.harvard.edu/#abs/2011ApJS..196...11S} {196, 11}

\bibitem[\protect\citeauthoryear{{Speagle}, {Steinhardt}, {Capak}  \&
  {Silverman}}{{Speagle} et~al.}{2014}]{Speagle2014}
{Speagle} J.~S.,  {Steinhardt} C.~L.,  {Capak} P.~L.,   {Silverman} J.~D.,
  2014, \mn@doi [\apjs] {10.1088/0067-0049/214/2/15}, \href
  {https://ui.adsabs.harvard.edu/abs/2014ApJS..214...15S} {214, 15}

\bibitem[\protect\citeauthoryear{{Tacchella}, {Dekel}, {Carollo}, {Ceverino},
  {DeGraf}, {Lapiner}, {Mand elker}  \& {Primack Joel}}{{Tacchella}
  et~al.}{2016}]{Tacchella2016}
{Tacchella} S.,  {Dekel} A.,  {Carollo} C.~M.,  {Ceverino} D.,  {DeGraf} C.,
  {Lapiner} S.,  {Mand elker} N.,   {Primack Joel} R.,  2016, \mn@doi [\mnras]
  {10.1093/mnras/stw131}, \href
  {https://ui.adsabs.harvard.edu/abs/2016MNRAS.457.2790T} {457, 2790}

\bibitem[\protect\citeauthoryear{{Tomczak} et~al.,}{{Tomczak}
  et~al.}{2016}]{Tomczak2016}
{Tomczak} A.~R.,  et~al., 2016, \mn@doi [\apj] {10.3847/0004-637X/817/2/118},
  \href {https://ui.adsabs.harvard.edu/abs/2016ApJ...817..118T} {817, 118}

\bibitem[\protect\citeauthoryear{{Wang} et~al.,}{{Wang}
  et~al.}{2011}]{Wang2011}
{Wang} J.,  et~al., 2011, \mn@doi [\mnras] {10.1111/j.1365-2966.2010.17962.x},
  \href {https://ui.adsabs.harvard.edu/\#abs/2011MNRAS.412.1081W} {412, 1081}

\bibitem[\protect\citeauthoryear{{Whitaker}, {van Dokkum}, {Brammer}  \&
  {Franx}}{{Whitaker} et~al.}{2012}]{Whitaker2012}
{Whitaker} K.~E.,  {van Dokkum} P.~G.,  {Brammer} G.,   {Franx} M.,  2012,
  \mn@doi [\apj] {10.1088/2041-8205/754/2/L29}, \href
  {https://ui.adsabs.harvard.edu/\#abs/2012ApJ...754L..29W} {754, L29}

\bibitem[\protect\citeauthoryear{{Whitaker} et~al.,}{{Whitaker}
  et~al.}{2014}]{Whitaker2014}
{Whitaker} K.~E.,  et~al., 2014, \mn@doi [\apj] {10.1088/0004-637X/795/2/104},
  \href {https://ui.adsabs.harvard.edu/abs/2014ApJ...795..104W} {795, 104}

\bibitem[\protect\citeauthoryear{{Whitaker} et~al.,}{{Whitaker}
  et~al.}{2015}]{Whitaker2015}
{Whitaker} K.~E.,  et~al., 2015, \mn@doi [\apjl] {10.1088/2041-8205/811/1/L12},
  \href {https://ui.adsabs.harvard.edu/abs/2015ApJ...811L..12W} {811, L12}

\bibitem[\protect\citeauthoryear{{Wright} et~al.,}{{Wright}
  et~al.}{2010}]{Wright2010}
{Wright} E.~L.,  et~al., 2010, \mn@doi [\aj] {10.1088/0004-6256/140/6/1868},
  \href {https://ui.adsabs.harvard.edu/\#abs/2010AJ....140.1868W} {140, 1868}

\bibitem[\protect\citeauthoryear{{Wuyts} et~al.,}{{Wuyts}
  et~al.}{2011}]{Wuyts2011}
{Wuyts} S.,  et~al., 2011, \mn@doi [\apj] {10.1088/0004-637X/742/2/96}, \href
  {https://ui.adsabs.harvard.edu/#abs/2011ApJ...742...96W} {742, 96}

\bibitem[\protect\citeauthoryear{{Zibetti}, {Charlot}  \& {Rix}}{{Zibetti}
  et~al.}{2009}]{Zibetti2009}
{Zibetti} S.,  {Charlot} S.,   {Rix} H.-W.,  2009, \mn@doi [\mnras]
  {10.1111/j.1365-2966.2009.15528.x}, \href
  {https://ui.adsabs.harvard.edu/#abs/2009MNRAS.400.1181Z} {400, 1181}

\bibitem[\protect\citeauthoryear{{Zolotov} et~al.,}{{Zolotov}
  et~al.}{2015}]{Zolotov2015}
{Zolotov} A.,  et~al., 2015, \mn@doi [\mnras] {10.1093/mnras/stv740}, \href
  {https://ui.adsabs.harvard.edu/abs/2015MNRAS.450.2327Z} {450, 2327}

\bibitem[\protect\citeauthoryear{{van der Wel} et~al.,}{{van der Wel}
  et~al.}{2012}]{vanderWel2012}
{van der Wel} A.,  et~al., 2012, \mn@doi [\apjs] {10.1088/0067-0049/203/2/24},
  \href {https://ui.adsabs.harvard.edu/abs/2012ApJS..203...24V} {203, 24}

\makeatother
\end{thebibliography}




\appendix

\end{document}